# Link between interlayer hybridization and ultrafast charge transfer in WS$_2$-graphene heterostructures

Niklas Hofmann[1], Leonard Weigl[1], Johannes Gradl[1], Neeraj Mishra[2,3], Giorgio Orlandini[2], Stiven Forti[2], Camilla Coletti[2,3], Simone Latini[4], Lede Xian[5], Angel Rubio[6], Dilan Perez Paredes[7], Raul Perea Causin[8], Samuel Brem[7], Ermin Malic[7,8], and Isabella Gierz[1]

[1] Department for Experimental and Applied Physics, University of Regensburg, Regensburg, Germany
[2] Center for Nanotechnology Innovation@NEST, Istituto Italiano di Tecnologia, Pisa, Italy
[3] Graphene Labs, Istituto Italiano di Tecnologia, Genova, Italy
[4] Department of Physics, Technical University of Denmark, Hovedstaden, Denmark
[5] Songshan Lake Materials Laboratory, Dongguan, China
[6] Theory Department, Max Planck Institute for the Structure and Dynamics of Matter, Hamburg, Germany
[7] Department of Physics, University of Marburg, Marburg, Germany
[8] Department of Physics, Chalmers University of Technology, Gothenburg, Sweden

E-mail: isabella.gierz@ur.de

## Abstract (max. 300 words)

Ultrafast charge separation after photoexcitation is a common phenomenon in various van-der-Waals (vdW) heterostructures with great relevance for future applications in light harvesting and detection. Theoretical understanding of this phenomenon converges towards a coherent mechanism through charge transfer states accompanied by energy dissipation into strongly coupled phonons. The detailed microscopic pathways are material specific as they sensitively depend on the band structures of the individual layers, the relative band alignment in the heterostructure, the twist angle between the layers, and interlayer interactions resulting in hybridization. We used time- and angle-resolved photoemission spectroscopy combined with tight binding and density functional theory electronic structure calculations to investigate ultrafast charge separation and recombination in WS$_2$-graphene vdW heterostructures. We identify several avoided crossings in the band structure and discuss their relevance for ultrafast charge transfer. We relate our own observations to existing theoretical models and propose a unified picture for ultrafast charge transfer in vdW heterostructures where band alignment and twist angle emerge as the most important control parameters.

## 1. Introduction

The great abundance of high-quality 2D materials with a broad range of electronic properties together with the ability to combine them into heterostructures with arbitrary stacking allows for the fabrication of novel artificial materials with tailored electronic properties. These van-der-Waals (vdW) heterostructures [1] not only present a way for realizing hitherto elusive exotic states of matter [2], they also bare great potential for future applications [3, 4, 5, 6, 7].

In the latter context, monolayer MX$_2$ (M=Mo,W and X=S,Se) has received particular attention. Monolayer MX$_2$ is a direct gap semiconductor with giant exciton binding energies and strong excitonic absorption in the visible spectral range. Further, due to strong spin-orbit coupling and broken inversion symmetry the valence band maximum shows a sizable spin splitting resulting in spin- and valley-selective optical excitation with circularly polarized light [8, 9, 10]. Graphene, on the other hand, posseses quite complementary properties: it exhibits a gapless conical dispersion, high carrier mobility, and negligible spin-orbit coupling resulting in long spin lifetimes [11].

Merging the properties of MX$_2$ and graphene in a vdW heterostructure thus holds great promise for emerging electronic properties with possible applications in the field of optoelectronics and optospintronics. Such heterostructures exhibit type I band alignment where both the valence band maximum (VBM) and the conduction band minimum (CBM) are located in the graphene layer. Hence, photodoping of WS$_2$ is followed by efficient charge transfer of both electrons and holes into the graphene layer [12]. Hole transfer is found to be much faster than electron transfer, resulting in a transient charge separated state [13, 14, 15]. In combination with cirularly polarized light this charge transfer might be exploited for optical spin injection into graphene [16].

Despite the huge potential for applications, the microscopic mechanism resulting in ultrafast charge separation in WS$_2$-graphene vdW heterostructures remains controversial [15, 17, 18, 19]. For this reason, tailoring the transfer rates of electrons and holes for selected applications remains challenging. Here, we address this issue with time- and angle-resolved photoemission spectroscopy (trARPES) that allows for a direct tracking of the relaxation pathways of photoexcited electrons and holes in the band structure. Tight binding and density functional theory electronic structure calculations reveal strong hybridization between the WS$_2$ and graphene layer in specific regions of the Brillouin zone. The resulting avoided crossings in the band structure and corresponding delocalization of the wavefunction act as efficient charge transfer channels. We relate our own observations to existing theoretical models and propose a unified picture for ultrafast charge transfer in vdW heterostructures where band alignment and twist angle emerge as the most important control parameters for the transfer rates.

## 2. Methods

### 2.4 sample growth

Pretreated 4H-SiC substrates (see SI for details) were H-etched to remove scratches and graphitized in Ar atmosphere. The resulting carbon monolayer with $(6\sqrt{3} \times 6\sqrt{3})R30°$ reconstruction was decoupled from the SiC substrate by H-intercalation yielding quasi-freestanding monolayer graphene [20]. WS$_2$ was then grown by chemical vapour deposition (CVD) from solid WO$_3$ and S precursors [21]. Both graphene and WS$_2$ growth were monitored using Raman spectroscopy, atomic force microscopy and secondary electron microscopy. WS$_2$ was found to grow in the shape of triangular islands with side lengths in the range of $300 - 700$nm with twist angles of either 0°(60°) or 30°(90°) with respect to the graphene layer.

### 2.2 tr-ARPES

The setup is based on a commercial titanium sapphire amplifier (Astrella, Coherent) with a central wavelength of 800nm, a repetition rate of 1kHz, a pulse duration of 35fs, and a pulse energy of 7mJ. 5mJ are used to seed a commercial optical parametric amplifier (Topas Twins, Light Conversion) the signal output of which is frequency doubled, yielding 2eV pump pulses resonant with the A-exciton of monolayer WS$_2$. The remaining 2mJ of output energy are frequency doubled and focused onto an Argon gas jet for high harmonics generation. A single harmonic at 21.7eV photon energy is selected with a grating monochromator yielding extreme ultraviolet (XUV) probe pulses that are used to eject photoelectrons from the sample. The photoelectrons are dispersed according to their kinetic energy and emission angle by a hemispherical analyzer (Phoibos 100, SPECS), yielding two-dimensional snapshots of the occupied part of the band structure in momentum space. The probe spot diameter was $\sim 250\mu m$ on the sample. Hence, the trARPES data represents an ensemble average of many different WS$_2$ islands. The energy and temporal resolutions for the measurements presented in the present publication are 200meV and 160fs, respectively.

### 2.3 tight binding

The band structure of the WS$_2$-graphene heterostructure is calculated with a microscopic model based on the second quantization formalism (see SI for details). We consider the Hamiltonian

$H = \sum_{l,\lambda,\mathbf{k}} E_{l,\lambda,\mathbf{k}} a^\dagger_{l,\lambda,\mathbf{k}} a_{l,\lambda,\mathbf{k}} + \sum_{l \neq l',\lambda,\mathbf{k},\mathbf{k}'} T^{ll'}_{\lambda \mathbf{k}\mathbf{k}'} a^\dagger_{l',\lambda,\mathbf{k}} a_{l,\lambda,\mathbf{k}}$

where the operator $a^{(\dagger)}_{l,\lambda,\mathbf{k}}$ annihilates (creates) an electron with momentum $\mathbf{k}$ at band $\lambda$ of layer $l$. Here, $E_{l,\lambda,\mathbf{k}}$ is the layer-specific electronic band structure, which is described with conventional tight binding models taking into account $p_z$ orbitals of carbon atoms in graphene [22] and $d_{z^2}$, $d_{xy}$, and $d_{x^2-y^2}$ orbitals of tungsten atoms in WS$_2$ [23]. The second term in the Hamiltonian describes interlayer tunneling with the matrix element $T^{ll'}_{\lambda k k'} = \langle \psi^{l'}_{\lambda k'} | V_t | \psi^l_{\lambda k} \rangle$, which is given by the overlap between wave functions $\psi^l_{\lambda k}$ of initial and final states and the tunneling potential $V_t$. In the tight binding approach, the tunneling matrix element is determined by orbital overlaps, which we approximate with Gaussian profiles considering realistic parameters (see SI). Next, we calculate the hybrid electron eigenstates in the Moiré mini-Brillouin zone by zone-folding and diagonalizing the Hamiltonian $H$ [24, 25]. A more clear representation of the hybrid band structure is obtained by unfolding the band structure [26] from the mini-Brillouin zone to the Brillouin zone of graphene.

*2.4 density functional theory*

We performed density functional theory (DFT) bandstructure calculations based on the local density approximation (LDA) functional for an untwisted WS$_2$-graphene heterostructure by employing a $4 \times 4$ and a $5 \times 5$ supercell for graphene and WS$_2$, respectively. This unit cell size allows us to minimize the amount of strain, while keeping the computational effort feasible. The calculations are perfomed in the GPAW code utilizing dzp LCAO basis functions [27, 28]. The bandstructure of the heterostructure was unfolded onto the graphene unit cell. The orbital character of the individual bands is established by integrating the probability density over the region of space belonging to either of the layers.

## 3. Results

Figure 1a shows a tr-ARPES snapshot of the band structure of the WS$_2$-graphene heterostructure along the ΓK-direction close to the WS$_2$ and graphene K-points. The snapshot was taken at negative pump-probe delay before the arrival of the pump pulse. White dashed lines are theoretical band structures for isolated WS$_2$ [29] and graphene monolayers [30], respectively, that were shifted in energy to match the experimentally observed band gap and doping level. Within our energy resolution of 200meV the experimental data is well described by a simple superposition of the band structures of the individual layers. The band marked by a red dashed line originates from WS$_2$ islands with a twist angle of 30°(90°) with respect to the graphene layer [29]. Figure 1b shows the pump-induced changes of the photocurrent with respect to Fig. 1a for a pump-probe delay of $t = 240$fs after photoexcitation of the heterostructure with 2eV pump pulses with a fluence of $F = 1.5$mJ/cm$^2$. Red and blue colours indicate gain and loss of photoelectrons, respectively. We observe a gain of photoelectrons at the bottom of the conduction band of WS$_2$ and a gain (loss) of photoelectrons in the graphene π-band above (below) the Fermi level. The strongest signal, however, is observed in the valence band region of WS$_2$ with a loss of photoelectrons at the equilibrium position of the spin-orbit split band and a gain above the equilibrium position.

To evaluate this complex pump-probe signal we start by extracting the time-dependent population of different regions in momentum space that are indicated by coloured boxes in Fig. 1b. The population dynamics of the graphene π-band above and below the Fermi level is shown in Fig. 2a together with single-exponential fits (see SI). We find that gain and loss are asymmetric with lifetimes of $\tau = 0.30 \pm 0.03$ps and $\tau = 2.1 \pm 0.3$ps, respectively. The gain of photoelectrons above the equilibrium position of the WS$_2$ valence band (Fig. 2b) shows an exponential lifetime of $\tau = 2.5 \pm 0.3$ps. The lifetime of photoexcited electrons at the bottom of the WS$_2$ conduction band (Fig. 2c) is found to be $\tau = 0.94 \pm 0.07$ps.

Next, we search for possible changes of the transient band structure by determining the binding energies for the WS$_2$ valence and conduction band and the graphene Dirac cone as a function of pump-probe delay. To this end, we extract energy and momentum distribution curves (EDCs and MDCs), respectively, along the coloured lines in Fig. 1a that we fit with an appropriate number of Gaussian (for EDCs) or Lorentzian peaks (for MDCs, see SI). The transient binding energies of the WS$_2$ valence and conduction band are shown in Fig. 3a. Subtracting the valence band position from the conduction band position yields the transient band gap shown in Fig. 3b together with a single-exponential fit. We find that the band gap decreases from the equilibrium value of 2.08eV [15] to 1.97eV at $t\sim500$fs with a lifetime of $\tau = 0.6 \pm 0.2$ps. Assuming that the band gap shrinks symmetrically about its center, we can subtract the contribution of the transient band gap renormalization from the transient peak positions in Fig. 2a. This yields the data in Fig. 2c where both the WS$_2$ valence and conduction band show a remaining up-shift of $\sim$90meV with an exponential lifetime of $\tau = 2 \pm 1$ps. In Fig. 3d we plot the transient position of the Dirac cone that we obtain by multiplying the transient position of the MDC peak with the slope of the Dirac cone. We find that the Dirac cone shifts down by $\sim$75meV with a lifetime of $\tau = 0.5 \pm 0.1$ps.

In summary, we find that (i) the population dynamics in graphene are highly asymmetric with a short-lived gain and a long-lived loss, (ii) there is a strong gain of photoelectrons above the equilibrium position of the WS$_2$ valence band with a lifetime of $\tau = 2.5 \pm 0.3$ps, (iii) the transient band gap of WS$_2$ reduces by 3.5% with a lifetime of $\tau = 0.6 \pm 0.2$ps, (iv) after subtracting the contribution of the transient band gap renormalization, all WS$_2$ states show a remaining upshift of $\sim$90eV with a lifetime of $\tau = 0.9 \pm 0.4$ps, and (v) the Dirac cone shifts down by $\sim$70eV with a lifetime of $\tau = 0.5 \pm 0.1$ps.

## 4. Discussion

The experimental results in Figs. 1-3 are in good agreement with previous trARPES data supporting the previous interpretation in terms of a transient charge separated state [13, 15]. There, photoexcitation of the A-exciton in $WS_2$ is followed by efficient charge transfer of both electrons and holes into the graphene layer. The fact that hole transfer is considerably faster than electron transfer then results in the formation of a charge-separated transient state with a lifetime on the order of ~1ps. The transient charge separation with electrons located in the conduction band of $WS_2$ and holes located in the valence band of graphene explains the asymmetric population dynamics of the Dirac cone in Fig. 2a as well as the band shifts in Figs. 3c and d. The gain of photoelectrons above the equilibrium position of the $WS_2$ valence band contains contributions of both the transient band gap renormalization and the transient charge separation.

The obvious driving force for ultrafast charge transfer in vdW heterostrustures is the band alignment and the fact that photoexcited electrons and holes tend to relax to the conduction band minimum (CBM) and valence band maximum (VBM) of the heterostructure, respectively. For type I band alignment (see Fig. 4a) both the CBM and VBM are located in the same layer. Provided that the other layer is photoexcited, both electrons and holes are expected to change layer during relaxation. For type II band alignment (see Fig. 4b) the CBM and VBM are located in different layers such that either electron or hole transfer is expected [31, 7].

Initially, it seemed suprising that charge transfer can occur on ultrafast femtosecond to picosecond time scales despite the weak vdW coupling between the layers. This issue was resolved by envoking a coherent transfer channel allowing electrons and holes to oscillate back and forth between the layers with frequencies on the order of $10^{13} - 10^{14}$Hz [32, 33, 17]. This coherent oscillation is made possible by momentum-dependent orbital hybridization resulting in wave functions that are delocalized over both participating layers [33, 34]. Electrons and holes then transfer into the neighbouring layer at constant energy, requiring energy dissipation into phonons to prevent back-transfer and to allow for carrier relaxation into the CBM or VBM [32, 33, 34, 35, 17].

At this level, this phonon-assisted coherent charge transfer model is quite general. It becomes material specific for several reasons: (1) The locations of CBM and VBM obviously depend on the band structures of the individual layers and their relative alignment in the heterostructure. (2) Charge transfer relies on orbtial hybridization between the individual layers that sensitively depends on interlayer distance and twist angle. (3) Charge transfer rates depend on the size of the respective tunneling matrix element, the available phase space, the electron-phonon coupling strength, and the phonon frequencies.

In order to apply this model to our $WS_2$-graphene heterostructure we start by calculating the band structure of the heterostructure within the tight-binding approximation and using DFT. The unfolded hybrid band structures are shown in Fig. 5. The bands are colour-coded to illustrate the degree of hybridization. The size of the data points is linked to the probability of finding an electron in that particular state. Overall, the two theories provide very similar results. Minor differences in the k-space region around $k \approx 1\text{Å}^{-1}$ in the lowest conduction band are attributed to the comparative simplicity of the tight binding description. The heterostructure exhibits type I band alignment with both the CBM and the VBM located inside the graphene layer. Hence, both electrons and holes originally generated by excitonic absorption inside the $WS_2$ layer are expected to transfer into the graphene layer. Charge transfer is expected to occur in those regions of the Brillouin zone where the wave function is delocalized over both layers. Our calculations reveal several avoided crossings with delocalized wave functions (red in Fig. 5) both in the conduction and the valence band of the heterostructure.

Photoexcitation at resonance to the A-exciton of $WS_2$ intially places electrons and holes at the CBM and VBM at the K-point of the $WS_2$ layer where the graphene contribution to the wave function is negligible. In order to transfer into the graphene layer both electrons and holes need to reach the closest avoided crossing in the band structure of the heterostructure. In Fig. 5b we see that, for electrons, the closest charge transfer state is located around $k \approx 1\text{Å}^{-1}$, separated from the CBM by an energy barrier of $\Delta E \approx 500$meV. For holes, we find $k \approx 1.49\text{Å}^{-1}$ and $\Delta E \approx 180$meV. Overcoming these barriers requires excess kinetic energy which explains the previously observed pump fluence dependence of the transfer rates [15]. According to time-dependent DFT and non-adiabatic molecular dynamics simulations in [17] the in-plane bond stretching $E_{2g}$ mode of graphene and the out-of-plane $A_{1g}$ mode of $WS_2$ provide the required energy dissipation for charge transfer from $WS_2$ to graphene.

To optimize charge transfer rates of a given heterostructure (i.e. a given combination of materials) for desired applications the phonon-assisted coherent charge transfer model described above suggests the following control parameters: (1) The twist angle between the layers will affect the hybridization between the layers with immediate impact on charge transfer states [36, 37, 38]. (2) The relative band alignment of the heterostructure can be controlled via perpendicular electric fields which yields control over the available phase space [17, 39, 40].

## 5. Conclusion

We combined trARPES mesaurements with tight binding and DFT band structure calculations to investigate the microscopic pathways for ultrafast charge separation and recombination in the heterostructure. Despite the weak vdW interaction between the layers we find indications for strong hybridization enabling ultrafast phonon-assisted coherent charge transfer. Based on these findings we propose to vary twist angle and band alignment to tailor electron and hole transfer rates for specific applications.

## Acknowledgements

The research leading to these results has received funding from the Deutsche Forschungsgemeinschaft through CRC 1277 (Project 314695032) and CRC 1038 (Project 223848855), and the European Union's Horizon 2020 research and innovation program under grant agreement no. 851280-ERC-2019-STG and 881603-Graphene Core3.

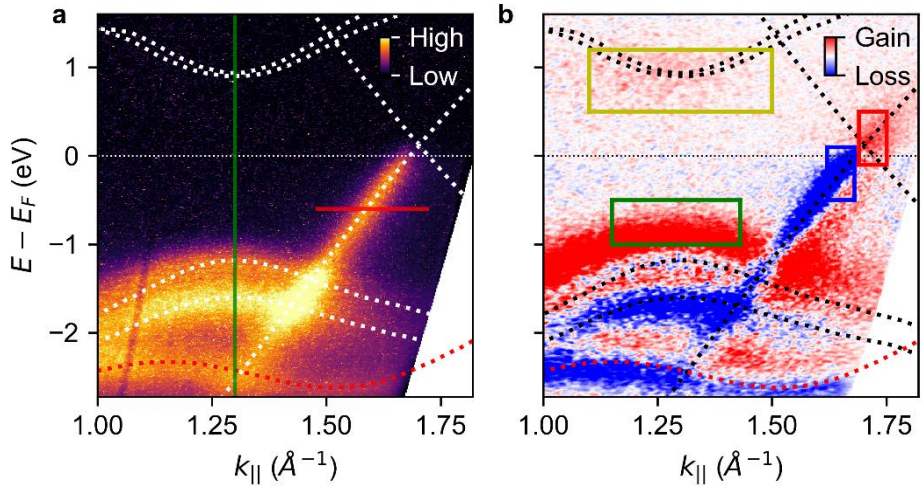

**Figure 1 | trARPES data of WS₂-graphene heterostructure. a)** Photocurrent along the ΓK-direction for negative pump-probe delay measured with p-polarized extreme ultraviolet pulses at 21.7eV photon energy. Dashed white lines mark the theoretical band structure of free-standing WS$_2$ [29] and graphene [30] monolayers, respectively, that were shifted in energy to match the experimentally observed band gap of 2.08eV [15] and the doping level of the Dirac cone of −100meV. The red dashed line marks the position of the WS$_2$ valence band for WS$_2$ islands with a twist angle of 30° with respect to the graphene layer. Continuous green and red lines indicate the positions of the line profiles used to extract the transient peak positions in Fig. 3. **b)** Pump-induced changes of the photocurrent 240fs after photoexcitation at a pump photon energy of 2eV with a pump fluence of 1.5mJ/cm². Gain and loss of photoelectrons are shown in red and blue, respectively. Coloured boxes indicate the area of integration for the pump-probe traces displayed in Fig. 2.

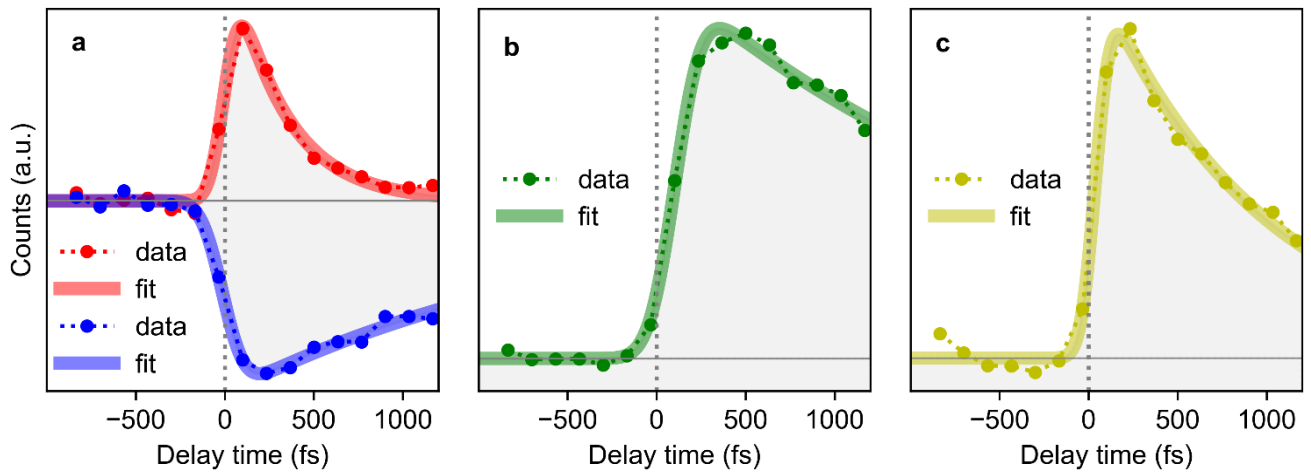

**Figure 2 | Population dynamics in different areas of momentum space.** The data points in this figure were obtained by integrating the photocurrent over the areas marked by the coloured boxes in Fig. 1b. The colour of the data points matches the colour of the corresponding box. Continuous thick lines are exponential fits to the data. **a)** Transient population of the Dirac cone of graphene above (red) and below (blue) the Fermi level. **b)** Gain signal above the equilibrium position of the WS$_2$ valence band. **c)** Transient population of the WS$_2$ conduction band.

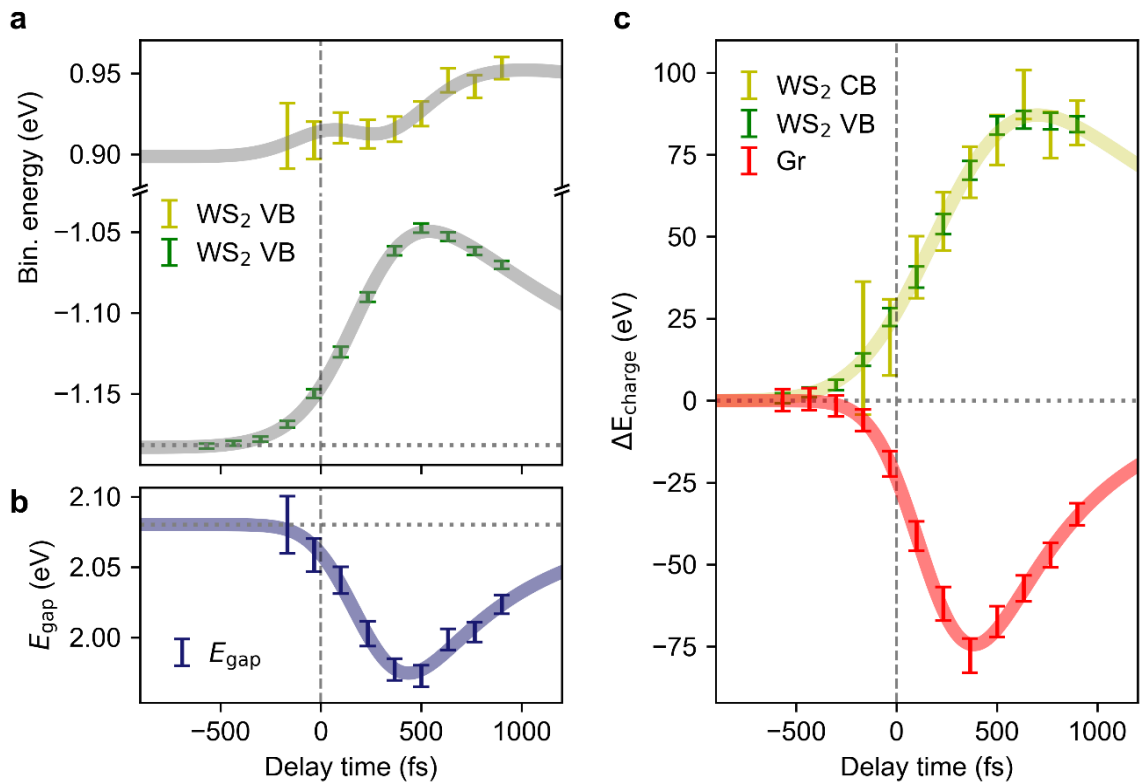

**Figure 3 | Transient band structure of WS$_2$-graphene heterostructure. a)** Transient positions of WS$_2$ valence (green) and conduction band (gold) at the K-point of WS$_2$. Thick grey lines are calculated from the exponential fits in b) and c). **b)** Transient band gap of WS$_2$ together with exponential fit. **c)** Band shifts due to charge separation of WS$_2$ (green and gold) and graphene (red) together with exponential fit.

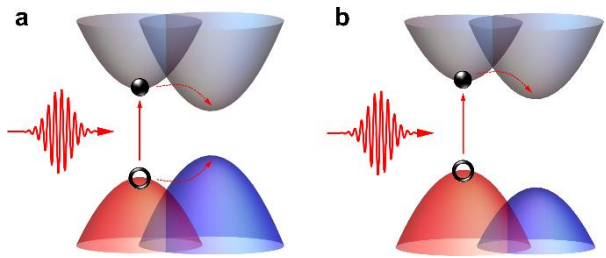

**Figure 4 | Influence of band alignment on ultrafast charge transfer. a)** Type I band alignment: resonant excitation of the layer with the bigger band gap (red) results in charge transfer of both electrons and holes into the layer with the smaller band gap (blue). The transfer rates for electrons and holes are, in general, not the same. **b)** Type II band alignment: photoexcitation of the red layer results in electron transfer into the blue layer. Similarly, excitation of the blue layer would result in hole transfer into the red layer.

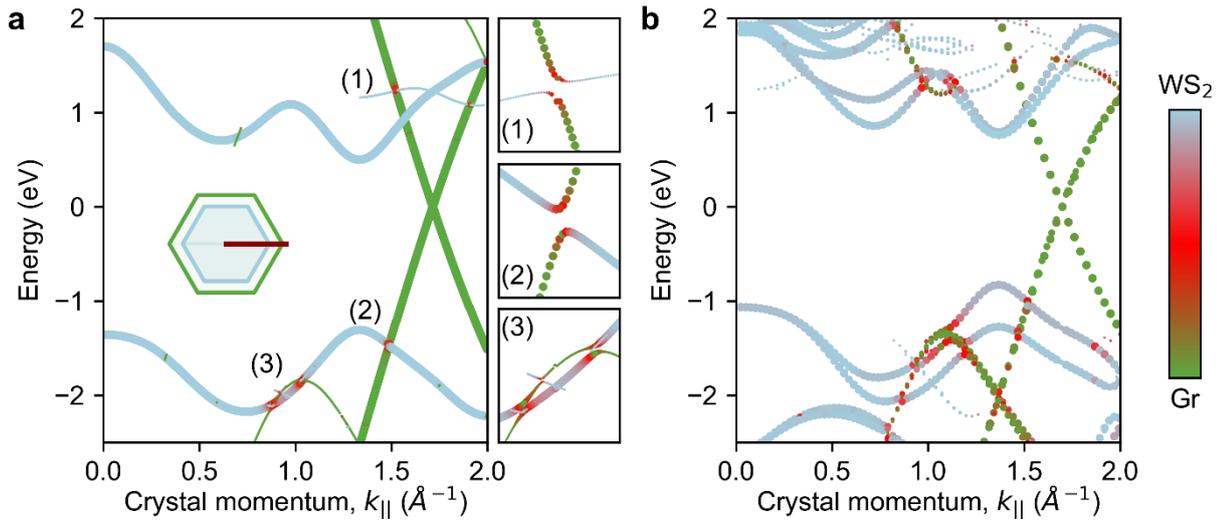

**Figure 5 | Band structure calculations for WS$_2$-graphene heterostructure with a twist angle of 0°. a)** Tight-binding model. **b)** DFT calculations. The size of the data points indicates the probability of finding an electron in that particular state. The colour code indicates the orbital composition of the wave function.